\documentclass[conference]{IEEEtran}
\usepackage{hyperref} 
\usepackage{amsmath, amssymb} 
\usepackage{courier} 
\usepackage{graphicx}
\graphicspath{ {images/} }
\usepackage{float} 
\usepackage{multirow} 

\usepackage{rotating}
\usepackage{subcaption}
\usepackage{url} 
\usepackage{rotating}
\usepackage{array} 
\usepackage[flushleft]{threeparttable}
\usepackage[utf8]{inputenc}
\usepackage{ulem}
\usepackage{color} 
\IEEEoverridecommandlockouts 

\begin{document}
\begin{titlepage}
\centering
	{\Large West Virginia University \par
	Lane Department of Computer Science and Electrical Engineering \par 
	Technical Report\par}
	\vspace{22\baselineskip}
	{\scshape\LARGE Malware Detection on General-Purpose Computers Using Power Consumption Monitoring: A Proof of Concept and Case Study \par}
	\vspace{2.5cm}
	\vfill
	\vspace{2cm}
	\vspace{2cm}
\textbf{\today} \\[2\baselineskip]
\end{titlepage}

\title{Malware Detection on General-Purpose Computers Using Power Consumption Monitoring: A Proof of Concept and Case Study 
\thanks{\tiny{This manuscript has been authored by UT-Battelle, LLC under Contract No. DE-AC05-00OR22725 with the U.S. Department of Energy. The United States Government retains and the publisher, by accepting the article for publication, acknowledges that the United States Government retains a non-exclusive, paid-up, irrevocable, world-wide license to publish or reproduce the published form of this manuscript, or allow others to do so, for United States Government purposes. The Department of Energy will provide public access to these results of federally sponsored research in accordance with the DOE Public Access Plan (http://energy.gov/downloads/doe-public-access-plan).}
}}

\author{
 \IEEEauthorblockN{Jarilyn M. Hern\'{a}ndez Jim\'{e}nez\IEEEauthorrefmark{1}\IEEEauthorrefmark{2}, 
 Jeffrey A. Nichols\IEEEauthorrefmark{1},
 Katerina Goseva-Popstojanova\IEEEauthorrefmark{2}, \\
 Stacy Prowell\IEEEauthorrefmark{1}, and Robert A. Bridges\IEEEauthorrefmark{1}} 
 \IEEEauthorblockA{\IEEEauthorrefmark{1} Computational Science and Engineering Division, Oak Ridge National Laboratory, Oak Ridge, TN 37831 \\
 \{hernandezjm1, nicholsja2, prowellsj, bridgesra\}@ornl.gov}
\IEEEauthorblockA{\IEEEauthorrefmark{2}Lane Department of Computer Science and Electrical Engineering, 
West Virginia University, WV,
Morgantown, 26506\\
\{jhernan7, katerina.goseva\}@mail.wvu.edu}} 

\maketitle

\begin{abstract}
Malware detection is challenging when faced with automatically generated and polymorphic malware, as well as with rootkits, which are exceptionally hard to detect. In an attempt to contribute towards addressing these challenges, we conducted a proof of concept study that explored the use of power consumption for detection of malware presence in a general-purpose computer. The results of our experiments indicate that malware indeed leaves a signal on the power consumption of a general-purpose computer. Specifically, for the case study based on two different rootkits, the data collected at the +12V rails on the motherboard showed the most noticeable increment of the power consumption after the computer was infected. Our future work includes experimenting with more malware examples and workloads, and developing data analytics approach for automatic malware detection based on power consumption.
\end{abstract}

 \section{Introduction}
\label{sec:intro}
Polymorphic malware can bypass signature-based detection methods and simple heuristic detection techniques by slightly changing the instructions of an existing malware sample. These new malware instances are called variants. Although these variants appear to be different programs from the viewpoint of signature-based anti-virus scanners, they exhibit similar functionality to their predecessor. Consequently, these new malware variants can bypass traditional detection methods until a signature for them can be identified and incorporated into detection software~\cite{2}.

Authors of malware detection systems have attempted to address this problem by using other methods that are more powerful than signature matching; for example, byte frequency~\cite{3}, general similarity measures~\cite{4}, and behavioral analysis~\cite{229} are among the proposed techniques. A common weakness of these detection methods is that they are executed on the same machine they are monitoring. Hence, successful attackers could disable the monitoring software or modify it to prevent detection after gaining entry to the system~\cite{7}. This behavior is evidenced by rootkits, a particularly insidious subclass of malware. Rootkits are a type of computer malware that were created to hide themselves and elude intrusion detection systems once they gain unauthorized access to a computer system~\cite{46}.

Previous work has also explored the idea of detecting the presence of malware by monitoring the power consumption of mobile devices, embedded systems, and software define radio. However, to the best of our knowledge, no one has explored if malware can be detected by monitoring the power consumption on general-purpose computers. 

Our goal in this paper is to prove the hypothesis that in order to mask themselves, rootkits will require a detectable change in the power consumption. Particularly, we are addressing the following research question: can we detect rootkits on general-purpose computers by analyzing only the power consumption? To this end we built a testbed and designed an experimental setup in which the power consumption  was recorded for a sequence of events running on a Windows operating system. 
This work focuses only on rootkits because they are commonly associated with the establishment of advanced persistent threats and pose serious danger to our nation's computer systems. Preliminary results showed that malware indeed leaves a signal on the power consumption of a general-purpose computer. Specifically, monitoring the +12V rails on the motherboard was the most useful for identifying the increase in the power consumption after the general-purpose computer was infected by malware.

The paper proceeds with related work in Section~\ref{sec:relWork}, followed by the experimental design in Section~\ref{sec:expdesign}, which includes the hardware and software setups used for collecting the power data, the experimental machine's execution of tasks, and  descriptions of the rootkit. Section~\ref{sec:results} presents the results of the feasibility study. Finally, conclusion and promising directions for future research are discussed in Section~\ref{sec:conclusions}.



\section{Related Work}
\label{sec:relWork}
Several works have used power consumption metrics for malware detection purposes. These methods have been tested on mobile devices~\cite{8,yang2016power}, embedded systems~\cite{9}, and software defined radio~\cite{10, gonzalez2014detecting, 250}. 

The work by Hoffman et al.~\cite{8} explored if malware can be detected on smartphones by analyzing their power consumption. This method failed due to the noise caused by unpredictable factors, such as user interaction and the mobile's signal strength. On the other side, the approach presented by Yang et al.~\cite{yang2016power} demonstrated that malware can be detected by monitoring the power consumption of smartphones. 
The difference between these two works is mainly in the type of smartphones used in the experiments. First method~\cite{8} focused mainly on ``old" devices (HTC-Nexus One and Samsung Galaxy Nexus), while the second method~\cite{yang2016power} focused on modern devices (Samsung Galaxy S5 and LG G2). 
Although PowerTutor~\cite{powerTutor} was used for the data collection in both works, this tool may have been updated between the time these two experiments were conducted, influencing the precision of the collected data and skewing the results.

Another method that monitors the power consumption on embedded systems with the objective of detecting malware was presented by Clark et al.~\cite{9}. Supervised machine learning techniques, such as 3-Nearest Neighbor, Multilayer Perceptron, and Random Forest, were used to analyze alternating current (AC) and to detect discrepancies among the power profiles. Even though the proposed approach share several similarities with this work, the main difference is that the work in~\cite{9} focused on monitoring the AC outlet, while we are monitoring several direct current (DC) channels. The problem with AC is that the current changes direction periodically, and because the current changes its direction the voltage reverses making the analog circuits much susceptible to noise.

Similarly, power-based malware detection for software defined radio was explored by Gonz\'{a}lez et al.~\cite{10, gonzalez2011}. This approach relied on extracting distinctive power consumption signatures and used pattern recognition techniques to determine if they matched the expected behaviors. This research was expanded, and used by the PFP firm (\url{http://pfpcyber.com}), which developed a commercial product that detect anomalies on a device by analyzing its power consumption. This approach can also be applicable to embedded devices~\cite{gonzalez2014detecting, 250}. The main difference between this approach and our work is that we monitor all the rails attached to the motherboard plus the CPU, while PFP is monitoring the power consumption of the device by placing a sensor on the processor's board as close to the power pins as possible.

It appears that there are no published research works focused on testing the use of power consumption monitoring in support of malware detection in general, with respect to the detection of rootkits in particular.

\section{Experimental Design and Data Collection}
\label{sec:expdesign}
The objective of our experiments is to analyze the power consumption of a general-purpose computer in order to detect the presence of rootkits. Our work is based on the hypothesis that rootkits can be detected by the anomalies they cause in the DC power consumption of the general-purpose computer. Specifically, we are interested in determining if there is a difference in the power profiles between the normal and 
anomalous behavior (i.e., after infection). 

\subsection{Hardware Configuration}
\label{hd_config}
Our experimental system is a Dell OptiPlex 755 with a clean installation of 32-bit Windows 7. The instrumentation for our experiments was a Data Acquisition system (DAQ), Model Number: USB-1608G Series \cite{234}. The DAQ connects to the device's motherboard power connector, and the voltage and current are collected on each of the DC power channels. The communication between this machine and the experimental machine was established through USB port. The DAQ provides relatively high-resolution power data, is able to sample at a rate of 250KHz, and can monitor up to 16 channels. Besides the DAQ, we also used an eight inch ATX power extender cable that had one male and one female 24-pin connector. The 24-pin male connector was attached to the motherboard, and the 24-pin female connector was attached to the power supply (PSU). 

Each group of wires on the PSU were connected to a single overcurrent protection (OCP) circuit that is called a \textit{rail}. A PSU has three voltage rails: \textit{+3.3V}, \textit{+5V}, and \textit{+12V}. Table~\ref{tab1} provides a list of the devices that are typically powered by these voltage rails. The +3.3V rails or the +5V rails are typically used by the digital electronic components and circuits in the system~\cite{99}. Some examples of these components are adapter cards and disk drive logic boards. On the other hand, the disk drive motors and the fans use the +12V rails~\cite{99}. Besides disk drive motors and newer CPU voltage regulators, the +12V supply is used by any cooling fans in the system~\cite {99}. 

\begin{center}
\begin{table}[H]
\caption{Voltage rail usage for a general-purpose computer}
\begin{tabular}{|c|c |} \hline
\bfseries{Rail} & \textbf{Devices Powered} \\ [0.5ex] 
 \hline
 +3.3V & chipsets, some DIMMs, PCI/AGP/PCIe cards, \\
 &miscellaneous chips  \\ \hline
 +5V & disk drive logic, low-voltage motors, SIMMs, \\
 &PCI/AGP/ISA cards, voltage regulators  \\ \hline
 +12V & motors, high-output voltage regulators, \\
 &AGP/PCIe cards  \\ \hline
 +12V CPU & CPU  \\ [1ex]
 \hline
\end{tabular}
\begin{tablenotes}
      \small
			\item Acronyms:
			\begin{itemize}
				\item SIMM = Single Inline Memory Module
				\item DIMM = Dual Inline Memory Module
				\item PCI = Peripheral Component Interconnect
				\item PCIe = PCI Express
				\item AGP = Accelerated Graphics Port
				\item ISA = Industry Standard Architecture
				\item CPU = Central Processing Unit
    \end{itemize}
		\end{tablenotes}
\label{tab1}
\end{table}
\end{center}

To ensure the power data was collected adequately, we tested three hardware configurations. The first hardware configuration monitored a total of eleven DC power channels (four pins had a signal of +3.3V,  five pins had a signal of +5V, and two pins had a signal of +12V). When using this configuration, the voltage levels were obtained for each of the channels that were monitored. However, since we were interested in power, both the voltage and current were required. To address this challenge, a DC voltage and current sense PCB was used.

The DC voltage and current sense PCB determines the DC current by measuring the voltage drop across a shunt resistor, and then converts that current to analog voltage output \cite{116}. 
The PCBs were soldered to those wires on the ATX power extender cable that we were interested in monitoring.

Leaving a total of thirteen DC power channels to be monitored (ten of them were used to measure the current and the other three channels were used to measure the voltage). Since the value of the voltage was the same, we measured the voltage as a group, that is, one voltage value for all the rails that were +3.3V, one value for all the rails that were +5V, and one value for all the rails that were +12V. While testing this hardware configuration, we noticed that there were two +12V rails that were powering the CPU of the experimental machine. Particularly, these +12V rails were separate from the rails that we were already monitoring on the ATX power extender cable. These +12V rails were connected from the PSU to a 4-pin ATX12V power connector on the motherboard. Including these rails, 
we ended up monitoring a total of fifteen channels. 
\footnote{A survey of other machines was made to verify that general-purpose computers have this 4-pin ATX12V power connector. More than 20 
computers were verified and all of them had the 4-pin ATX12V power connector.} 

\begin{figure}[H]
  \begin{minipage}[b]{0.47\linewidth}
  \centering
			\includegraphics[width=\textwidth]{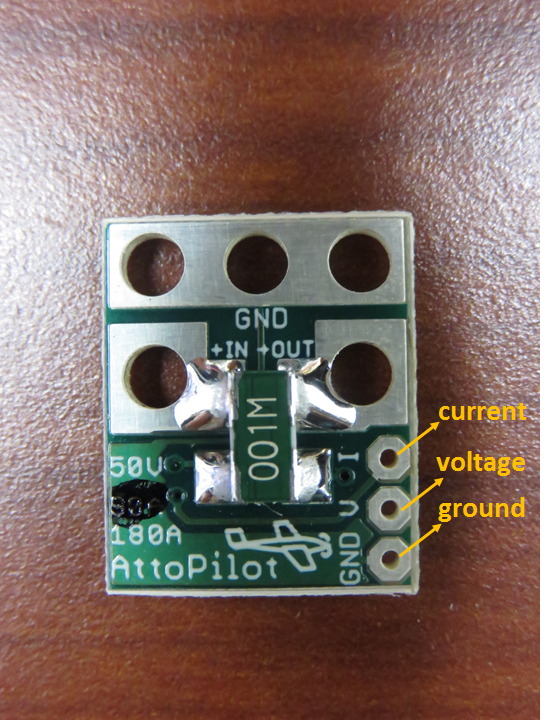}
  \end{minipage}
  ~\hfill~
  \begin{minipage}[b]{0.47\linewidth}
    \centering
		  \includegraphics[width=\textwidth]{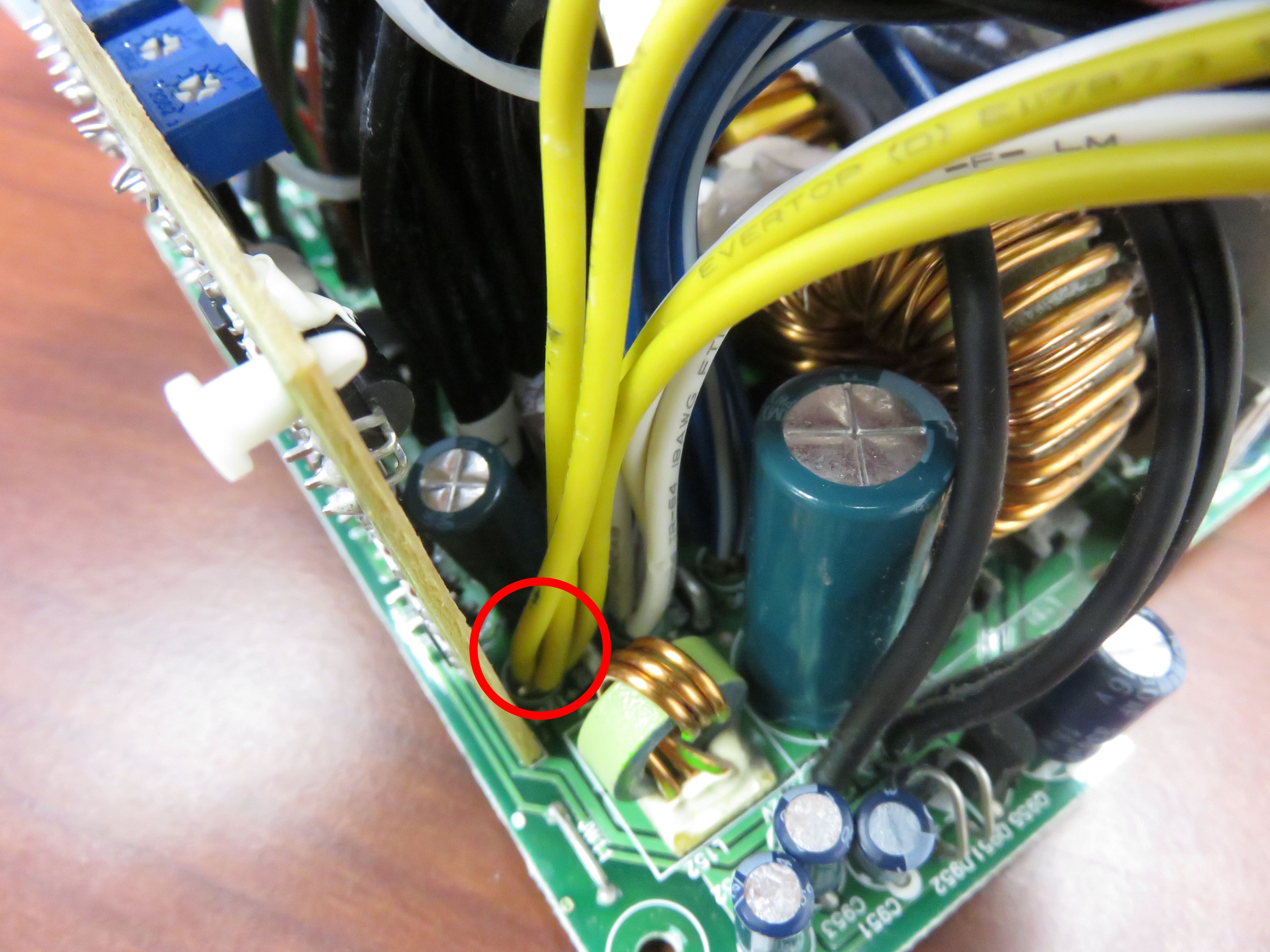}
\end{minipage}
\caption{(Left) Voltage and current sensor PCB used on the experiments for the second hardware configuration (Right) +12V rails soldered on the same contact point on the PSU.}
\label{fig:HW}
\end{figure}

Monitoring fifteen channels at the same time was challenging because, when post-processing, we had to sum several measured currents together. To simplify the hardware configuration, we evaluated other options that could help us to reduce the number of channels to be monitored. After some exploring we found that all wires from the same voltage value were soldered together on the same contact point on the power supply. This means that all the +3.3V rails were connected to the same contact point, and the same was true for the +5V rails, and the +12V rails. Figure~\ref{fig:HW}(left) shows the voltage and current sense PCB that was used for the second hardware configuration, while Figure~\ref{fig:HW}(right) shows how the +12V rails were soldered together on the same contact point on the PSU.

\begin{figure}[H]
  \begin{minipage}[b]{0.47\linewidth}
  \centering
			\includegraphics[width=\textwidth, angle=90]{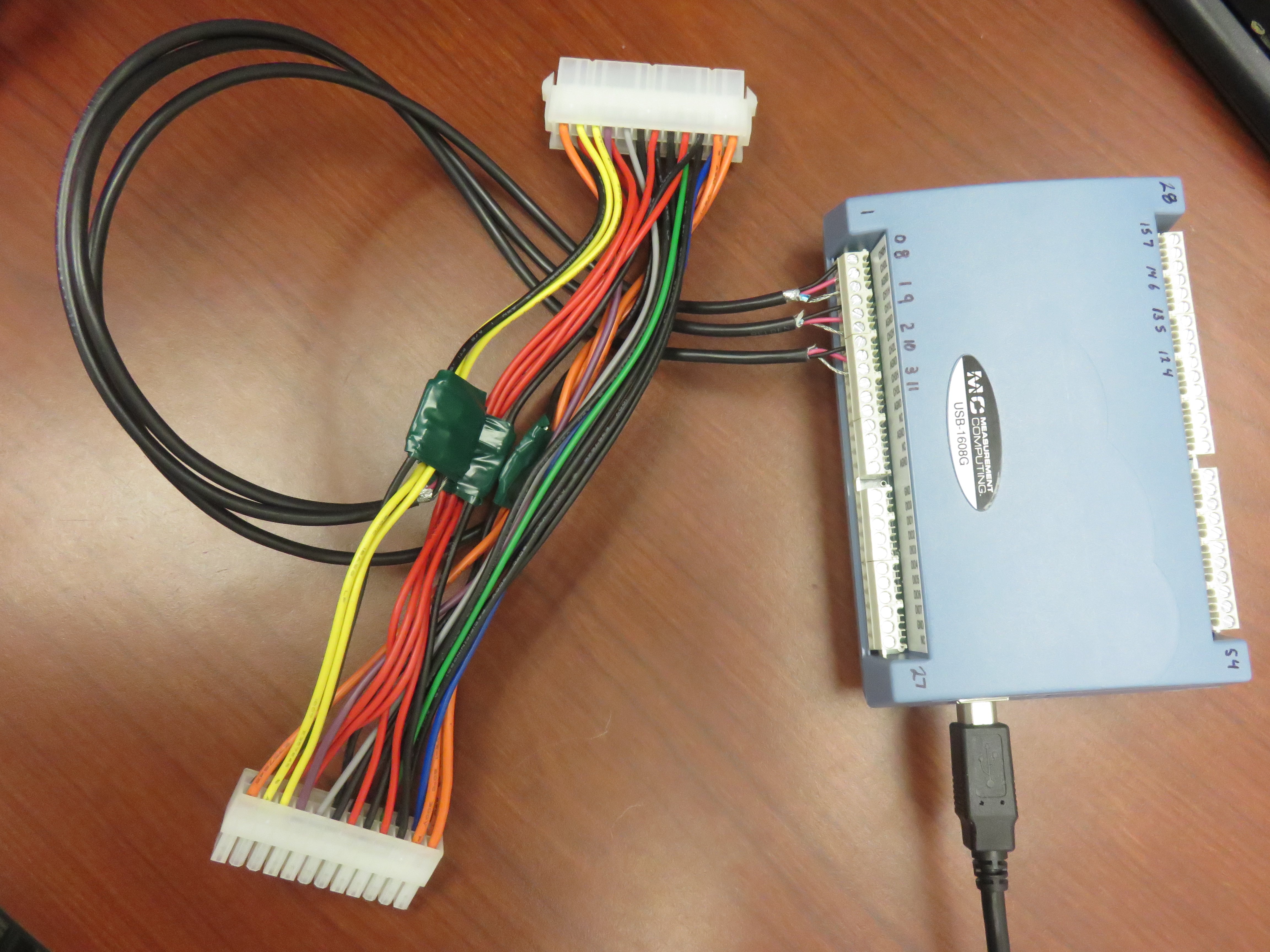}
  \end{minipage}
  ~\hfill~
  \begin{minipage}[b]{0.47\linewidth}
    \centering
		  \includegraphics[width=\textwidth]{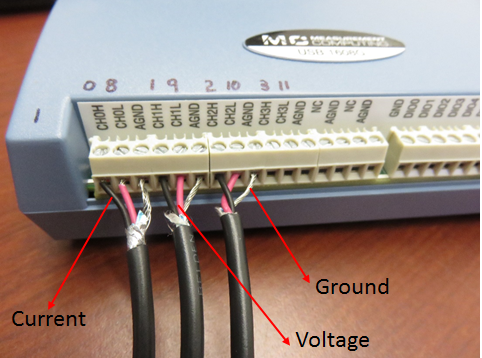}
\end{minipage}
\caption{(Left) Third hardware configuration used during the experiments (Right)  Wires attached to DAQ.}
\label{fig:HW2}
\end{figure}

The third hardware configuration emerged from this observation. We grouped all the +3.3V rails on the same voltage and current sense PCB which was attached to the ATX power extender cable; the same was done for the +5V rails and the +12V rails. Figure \ref{fig:HW2}(left) shows the third hardware configuration, while Figure~\ref{fig:HW2}(right) shows how the wires from the ATX power extender cable were hooked to the DAQ. As can be seen from Figure~\ref{fig:HW2}(right), for each channel on the DAQ we hooked a wire for the current (black wire), the voltage (red wire), and the ground (silver wire). 

Grouping the rails reduced the numbers of channels to be monitored to six\textemdash three channels for measuring the current, and the other three channels for measuring the voltage. In addition, we also included the two +12V rails that  power the CPU. Overall, instead of monitoring fifteen channels, we reduced the number to eight: 4 voltage channels and 4 corresponding current channels. This configuration was the one used for the experiments and data collection described here. 

\subsection{Software Configuration}
\label{soft_config}
Initially, 
we used a tool called \textit{TracerDAQ Pro} (version 2.3.1.0), which is an out-of-the box virtual instrument that acquires and displays data~\cite{234}. This tool ran on a different machine (data repository) in order to provide integrity during the experimentation process. The acquired data from the experimental machine was stored as a CSV file on the data repository. 

As our experimental design evolved, we found that TracerDAQ Pro was not suitable for  obtaining precise power data. To address this issue we developed our own Visual Basic program. There were three advantages to using our own software versus using the supplied software: (1) data has 16 bits of precision; 
(2) we have control over the sample recording rates and sample timing (3) we were able to make additional real-time calculations that helped us to verify the obtained power data.


Another application used in our experiments is called \textit{Clonezilla}~\cite{203}. Clonezilla is a partition and disk imaging/cloning program. This tool was used to ensure we had a consistent, clean installation of Windows. We used Clonezilla to create an exact copy of the master hard drive and this hard drive was not exposed to malware. 

\subsection{Data Collection}
\label{datacol}
The power consumption of the general-purpose computer was collected in two different scenarios: normal behavior (no rootkit running on the system) and anomalous behavior (a rootkit was running on the system). 
For the data collection workflow, we assumed a clean installation of Windows, then power data was collected and labeled as normal. Subsequently, the experimental machine was infected and power data was collected and labeled as anomalous. For this case study we infected the general-purpose computer with two rootkits: Alureon and Pihar. 

A segregated network was created to ensure the malware will not spread around the main network. The segregated network consisted of an experimental machine, a data collection repository, a hub, and a cellular data connection. The data collection repository connects to the personal hotspot, and then through the hub we shared the wireless connection with the experimental machine. Two advantages from the use of segregated network are: (1) allowing rootkits to behave normally, while avoiding the possibility of infecting other machines on the network; and (2) allowing us to monitor, record, and analyze the experimental machine's network traffic.

Wireshark was used to collect the network traffic of the experimental machine and to validate that the experimental machine was successfully infected with the rootkits being tested. As part of the network traffic analysis, we organized the protocols on the PCAP file by alphabetical order and then focused only on the column for the Domain Name System (DNS) protocol. From the domains that were captured, one of them got our attention (\textit{term0l5ter12.com}). Several websites \cite{243, 245} had this domain registered as malicious. After all these analyses, we  were certain that the experimental machine was successfully infected with both rootkits.

To initialize the data collection process, we wrote two scripts: a Python script that executes a sequence of events, and a C++ program that inserted what we called a \textit{marker}. The objective of the Python script was to ensure repeatability, while the objective of the marker was to insert a signal into the measured power data to mark the start and end points for each of the events.  
IE was chosen because the Alureon and Pihar rootkits affect the performance of browsers~\cite{117, 118}.

\begin{figure}[H]
\includegraphics[scale=0.30]{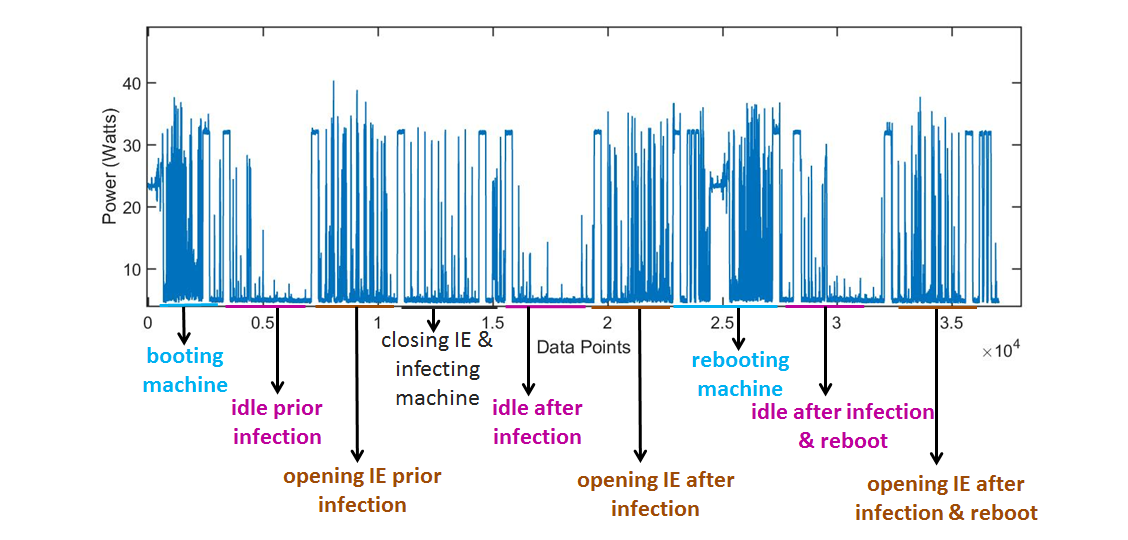}
\centering
\caption{Sequence of events after the Python script was executed}
\label{fig:SeqOfEvents}
\end{figure}

When the Python script is executed, it launches two markers before the experimental machine goes idle for a minute. Then the Python script opens ten windows of IE each with 5 seconds delay. Figure~\ref{fig:SeqOfEvents} shows data collected after the Python script was executed for the +12V CPU rail prior and after the infection with the Alureon rootkit. These events (idle, opening IE, booting/rebooting) were recorded during three states: (1) prior to infection, (2) after infection, and (3) after infection plus reboot. In order to segment these sections of the power profile, we used the marker to stress the CPU of the experimental machine for five seconds. The Python script places markers in the power data before and after the events were recorded. The advantage of using these markers is that they allow us to understand when a particular event occurs and how long it takes to complete its execution. This workflow was completed three times for the four rails that were monitored. 

The first rootkit, Alureon, also known as \textit{TDL4} or \textit{TDSS}, is a Trojan that allows an attacker to intercept incoming and outgoing Internet traffic in order to gather confidential information such as user names, passwords, and credit card data \cite{89}. There are several generations of this type of malware, and for our experiments, we used the fourth generation \cite{204}. Typically, it infects a computer via drive-by download through a questionable website, often a distributor of pornography or pirated media \cite{90}. Once Alureon is installed on the machine, the software searches the system for any competitor's malware and removes it. It also uses an encryption algorithm to hide its communications from traffic analysis tools that are sometimes used to detect suspicious transmissions \cite{90}. Furthermore, this rootkit can manipulate the master boot record (MBR) of the computer to ensure that it is loaded early during the bootup process so that it can interfere with the loading of the operating system \cite{207}. The second rootkit, which is a variant of Alureon, is a Trojan called \textit{Purple Haze} (also known as \textit{Pihar}). Like Alureon, this rootkit can modify the MBR of the machine, as well as changing system settings and reconfiguring the Windows registry. Its rootkit capabilities include disabling the antivirus software to keep itself hidden~\cite{ref:Pihar1}.


\subsection{Data Pre-processing}
\label{datepreprocessing}
As part of the data pre-processing, the voltage and current for the monitored rails were multiplied to obtain the power consumption of the general-purpose computer. To plot and interpret the power data, we used MATLAB. After obtaining the power data, the next step was to separate the events based on the start and end point.

To obtain the indexes we wrote a MATLAB script that returns the start and end point of all the markers that appeared on the dataset. For this case study, there are a total of eighteen markers. Once we had the start and end point for each event, the next step was to compare those events that were related to each other. Specifically, we were interested in the following comparisons: (1) when the machine was booting prior to infection versus when the machine was rebooting after infection; (2) idle prior to infection versus idle after infection; (3) idle prior to infection versus idle after infection and reboot; (4) when opening IE windows prior to infection versus when opening IE windows after infection (5) when opening IE windows prior to infection versus when opening IE windows after infection and reboot.

\section{Data Analysis and Case Study Results}
\label{sec:results}
The primary goal of this proof of concept is to determine if there is a difference in the power consumption of a general-purpose computer after malware infection. To prove or disprove this hypothesis, several experiments were conducted and power profiles were collected for specific events (idle, opening IE, and booting/rebooting). This was done for the rootkits Alureon and Pihar. For each rootkit there were three datasets. Each dataset contains the power consumption obtained for each of the rails that were monitored. In other words, each dataset contains the power profiles of all the sequence of events that were recorded for each one of the rails. The comparison between the normal and anomalous state was done for each of the events that were recorded on the four rails. Five graphs were generated for each monitored rail. 
The x axis for each of these graphs shows ``Data Points", which refers to the total of power readings that were sampled every 10 milliseconds. For example, if a graph shows 3,000 data points that would be equivalent to 30 seconds. 

\subsection{+3.3V Rails}
\label{3VRails}
These rails are typically used by digital electronic components and circuits in the system, such as memory. When comparing the power profiles of booting prior to infection versus when it was rebooting after infection, we noticed that at the beginning the power consumption was lower and subsequently both events kept their power consumption similar to each other. Regarding the other events (idle and opening IE), results showed that the difference in the power consumption cannot be established by the naked eye. After analyzing all six datasets (three datasets per rootkit), we concluded that the +3.3V rails are not very useful for detecting different behaviors between the normal and anomalous power profiles because these rails are used to power up memory, and that component does not consume as much power as the hard drive or CPU. 

\subsection{+5V Rails}
\label{5VRails}
For all datasets, when comparing booting prior to infection with the rebooting after infection for +5V rails , we noticed the same behavior as the +3.3V rails, that is the power consumption after infection was lower at the beginning of the initialization process, but later it kept the same pace as the normal behavior. Hence, comparing booting prior to infection versus booting after infection for the +5V rails is not sufficient to distinguish between normal and anomalous behavior. 

\begin{figure}[H]
\includegraphics[scale=0.15]{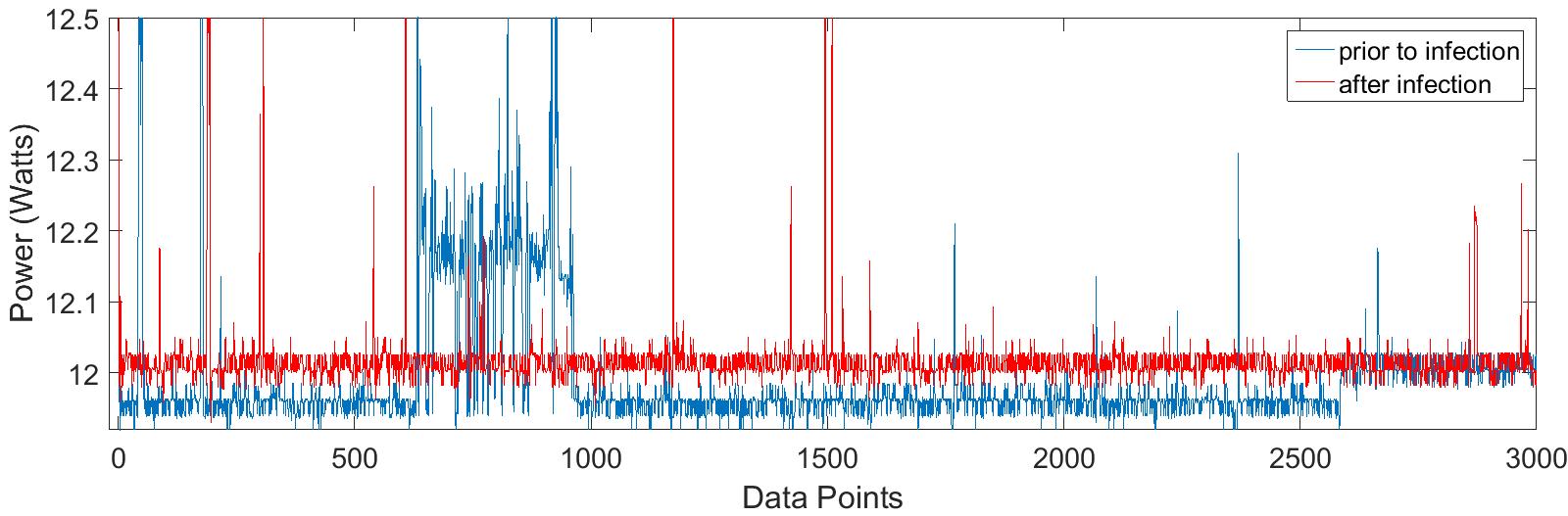}
\centering
\caption{Power consumption for idle prior to infection vs. idle after infection with Alureon for the +5V rails}
\label{fig:IdlePriorAfterInfection}
\end{figure}

When we compared idle prior to infection versus idle after infection with Alureon we obtained an increment in the power consumption after the general-purpose computer was infected for two out of the three datasets (66.67\% of the time), while for Pihar we noticed an increment in the power consumption for all datasets (100\% of the time). However, when comparing idle prior to infection versus idle after infection and reboot for both rootkits, we noticed that the power profiles for both scenarios (normal and anomalous) were at the same level. In other words, a distinguishable difference cannot be made by the naked eye. Furthermore, when comparing all the graphs in which the general-purpose computer was idle we noticed a delay in the power data after the general-purpose computer was infected. We believe this delay is because after the infection more processes are running and this extra work consumes more power. Figure~\ref{fig:IdlePriorAfterInfection} shows the power consumption after infecting the general-purpose computer with the Alureon rootkit. As can be seen from Figure~\ref{fig:IdlePriorAfterInfection}, the power consumption in the idle state was higher after the infection than prior to infection. Hence, this comparison is a good criterion for detecting malware through the power consumption.

\begin{figure}[H]
\includegraphics[scale=0.15]{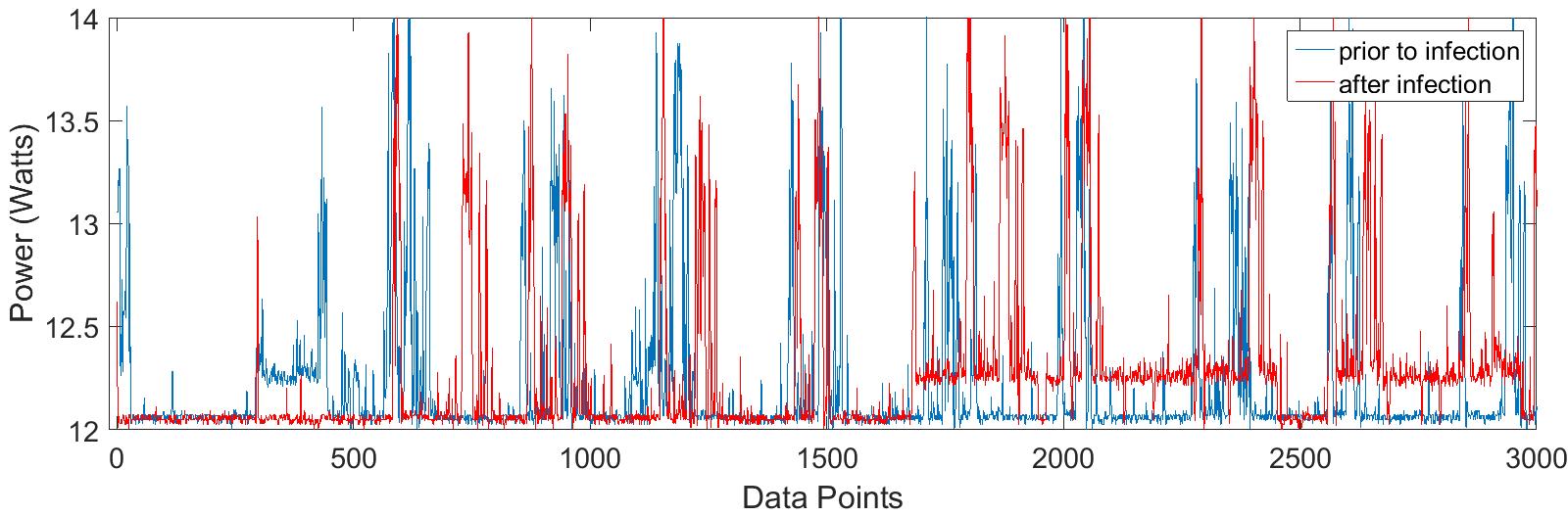}
\centering
\caption{Power consumption for opening IE prior to infection vs. opening IE after infection with Alureon for the +5V rails}
\label{fig:IEPriorAfterInfection}
\end{figure}

When IE was opened prior to infection versus after the infection with Alureon, we noticed an increment in the power consumption after infection for two out of three datasets (66.67\% of the time). In the case of the Pihar rootkit, this behavior was seen only in one out of three datasets (33.33\% of the time). Figure~\ref{fig:IEPriorAfterInfection} shows the power consumption after opening IE prior to infection versus after infection. From Figure~\ref{fig:IEPriorAfterInfection} we can see an increment in the power consumption when some IE windows were opened. Interestingly, this increment was seen when some windows of IE were jammed. This was consistent with the behavior we saw during the data collection process and later was confirmed when analyzing the PCAP file. Based on network traffic collected by Wireshark, we noticed that Alureon was trying to redirect the search engine to advertisement websites. However, when IE was opened prior to infection versus after the infection and reboot for both rootkits, a difference by the naked eye could not be established.

\subsection{+12V Rails on the Motherboard}
\label{12VRails}
\begin{figure}[H]
\includegraphics[scale=0.15]{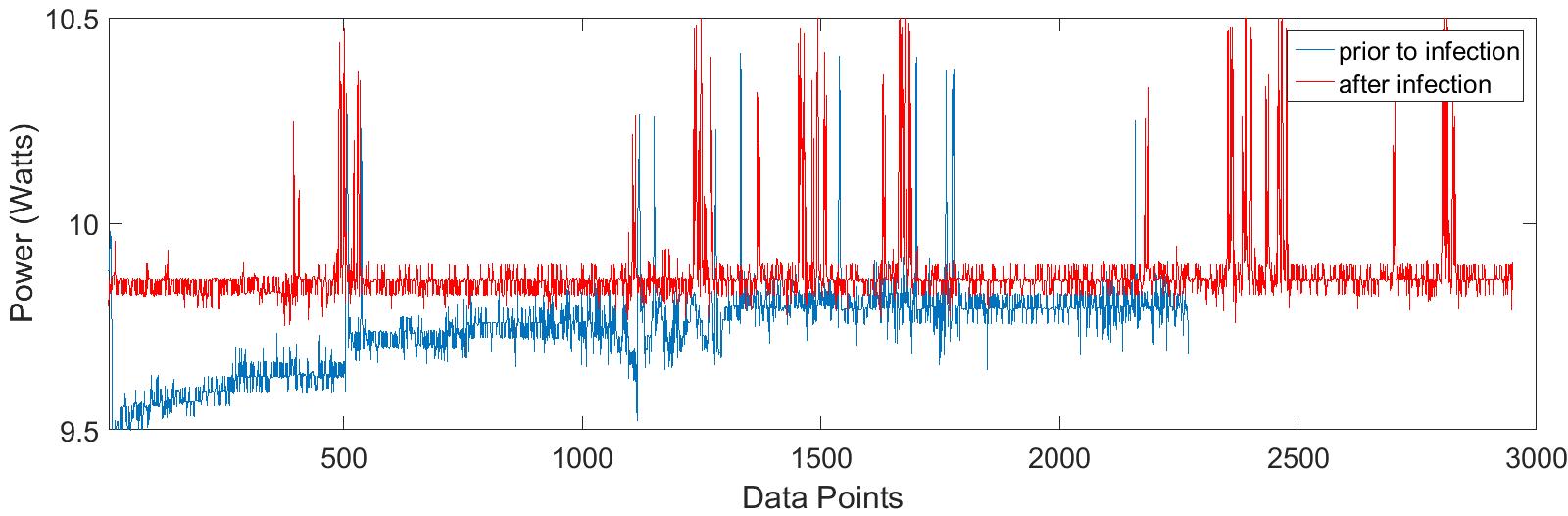}
\centering 
\caption{Power consumption for booting prior to infection vs. booting after infection with Alureon for the +12V rails on the motherboard}
\label{fig:BootingPriorAfterInfection}
\end{figure}

The +12V rails on the motherboard are used to power up the disk drive motors and the fans. For one of the Alureon datasets results showed that the power consumption was higher after the infection compared to when it was booted prior to infection (33.33\% of the time). However for the other two datasets, we saw similar behavior as in the case of +3.3V and +5V rails. Figure~\ref{fig:BootingPriorAfterInfection} shows an increment in the power consumption after the general-purpose computer was infected during the initialization process. In the case of Pihar, an increment in the power consumption was noticeable on two out of three datasets (66.67\% of the time).

\begin{figure}[H]
\includegraphics[scale=0.15]{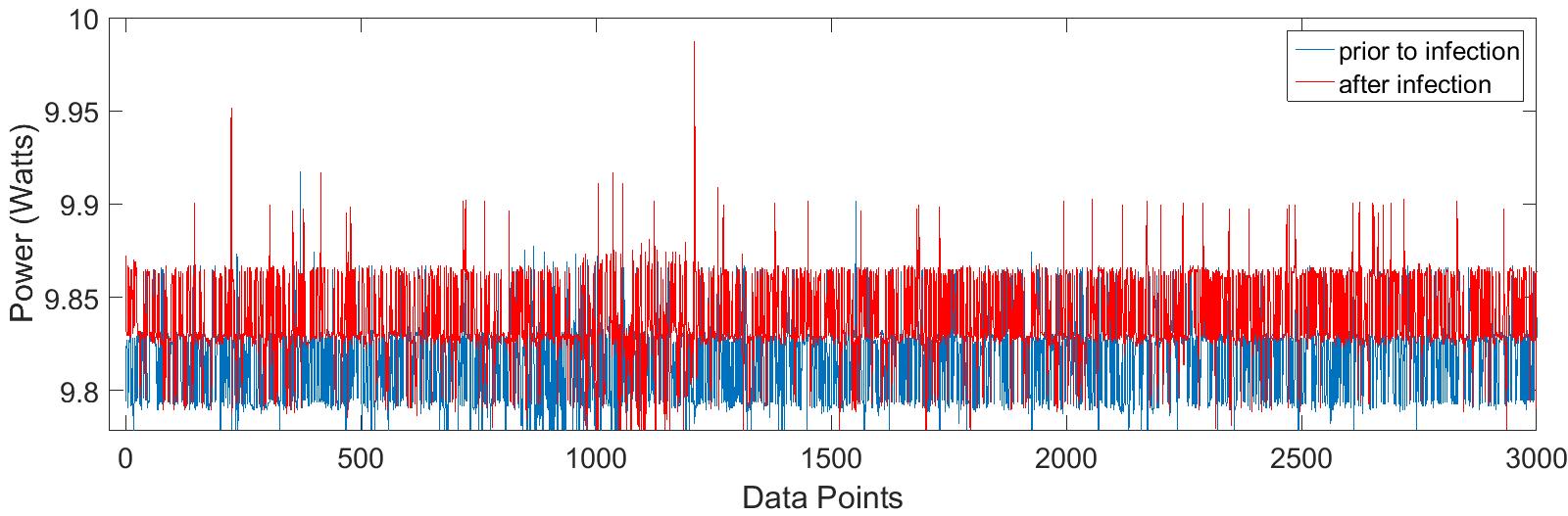}
\centering
\caption{Power consumption for idle prior to infection vs. idle after infection and reboot with Alureon for the +12V rails on the motherboard}
\label{fig:IdlePriorAfterInfectionAndReboot}
\end{figure}

When comparing the idle state (idle prior to infection versus idle after infection and idle prior to infection versus idle after infection and reboot), results for Alureon showed an increment in the power consumption after infection for two out of the three datasets (66.67\% of the time). Similar increment was seen in all three datasets of Pihar (100\% of the time). Figure~\ref{fig:IdlePriorAfterInfectionAndReboot} shows an increment in the power consumption when comparing idle prior to infection versus idle after infection and reboot for the Alureon rootkit. 

\begin{figure}[H]
\includegraphics[scale=0.15]{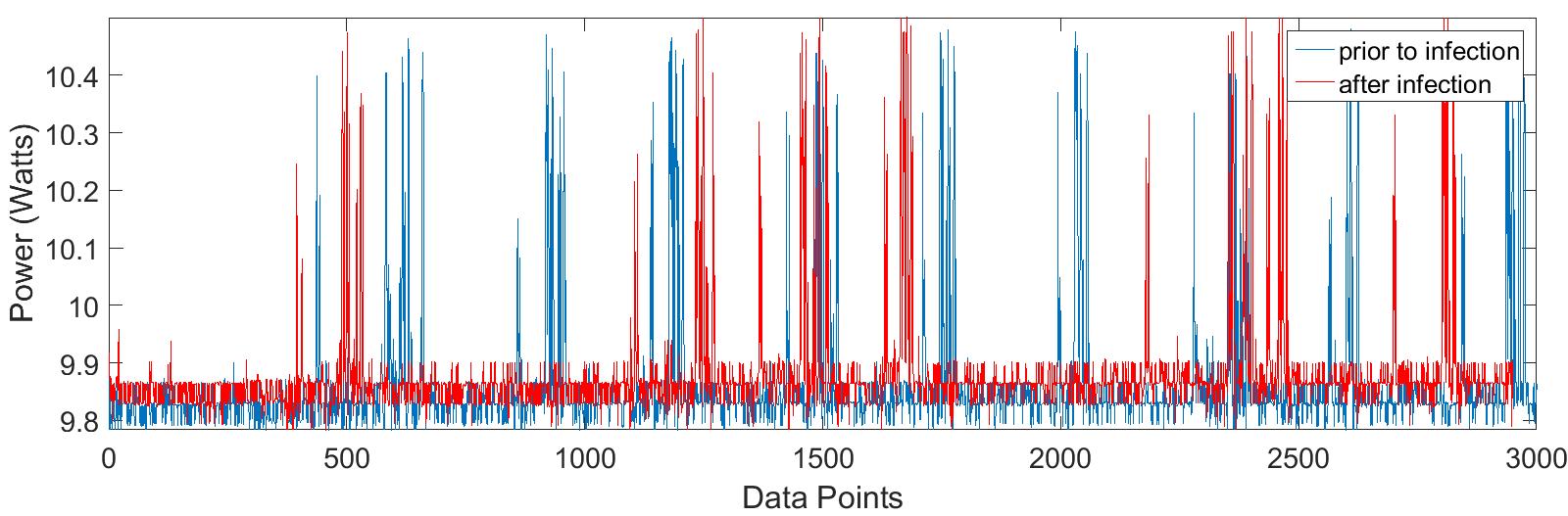}
\centering
\caption{Power consumption for opening IE prior to infection vs. opening IE after infection and reboot with Alureon for the +12V rails on the motherboard}
\label{fig:IEPriorAfterInfectionAndReboot}
\end{figure}

Nonetheless, when comparing IE (IE prior to infection versus after infection and IE prior to infection versus after infection and reboot), results for Alureon showed that an increment in the power consumption after infection can be seen in only one of the datasets (33.33\% of the time). Figure~\ref{fig:IEPriorAfterInfectionAndReboot} shows an increment in the power consumption when comparing IE prior to infection versus IE after the Alureon infection and reboot. After analyzing the +12V rails on the motherboard, we concluded these rails are very useful when analyzing the normal and anomalous power profiles.

When comparing IE prior to infection versus IE after infection for Pihar, we noticed an increment in the power consumption after infection for one out of the three datasets (33.33\% of the time). Interestingly, when comparing IE prior to infection versus IE after infection and reboot we noticed the power consumption of the general-purpose computer was higher after infection for all datasets (100\% of the time).

\subsection{+12V CPU Rails}
\label{12VCPURail}
The +12V CPU rails are separate from the +12V rails on the motherboard (monitored in the PSU). They are used to power the CPU or GPU of a general-purpose computer. 
The +12V rails on the motherboard are used to power disk drive motors and fans

The comparison between the power consumption when the general-purpose computer was booting prior to infection versus when it was booting after infection showed that at the beginning of the initialization process  the power consumption was higher prior to infection for both rootkits. However, at some point during the initialization, an increment in the power consumption after infection was noticeable. This comparison by itself does not provide information that can help us to distinguish between normal and anomalous behavior because of the presence of noise. Noise is expected during the booting and rebooting process because the system is executing several processes simultaneously, so even if the malware is present, its challenging to differentiate between normal and anomalous states. 

In the case of idle (idle prior to infection versus after infection and idle prior to infection versus after infection and reboot), we noticed that the power consumption for both rootkits in the normal and anomalous scenarios were similar. However, there were some higher spikes after infection. We believe these spikes were generated when the system was executing normal ``non malicious processes". Similarly, these spikes were also seen in the +5V rails. To be sure about the cause of these spikes, as part of our future work, we plan to collect other parameters such as kernel events, registry files, or syslogs of the general-purpose computer and correlate this information with the power consumption.

\begin{figure}[H]
\includegraphics[scale=0.15]{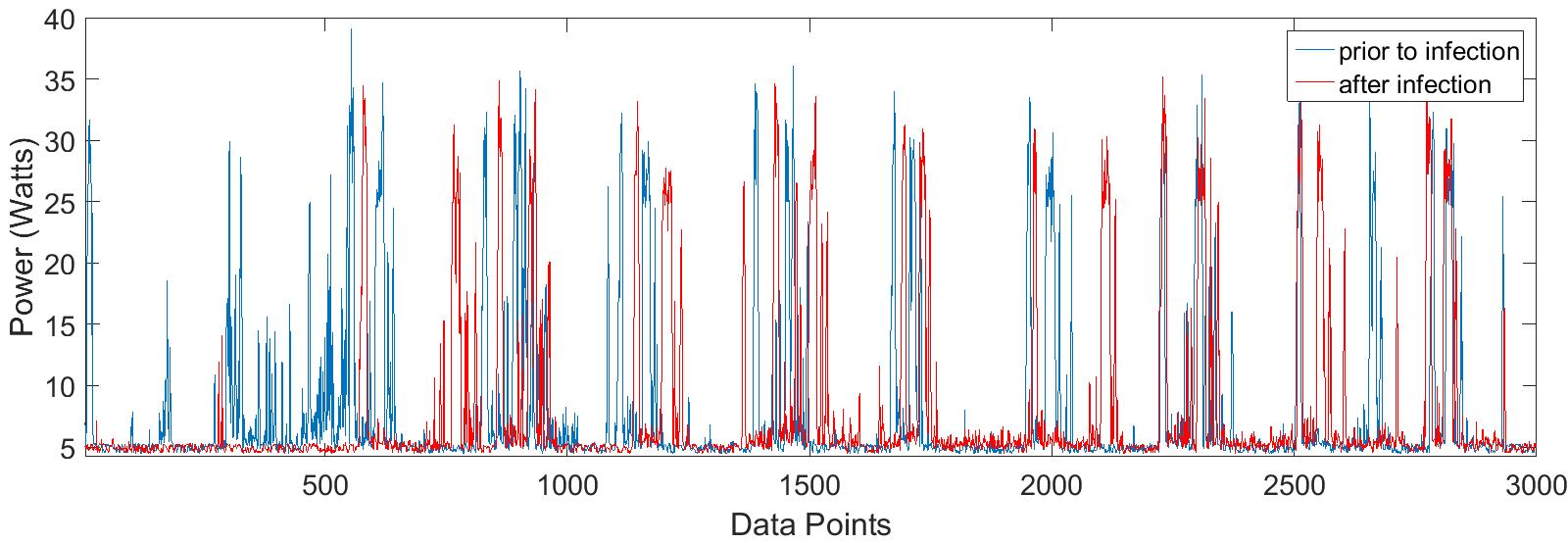}
\centering
\caption{Power consumption for opening IE prior to infection vs. opening IE after infection with Alureon for the +12V CPU rails}
\label{fig:IEPriorAfterInfection12VCPU}
\end{figure}

A similar behavior was noticeable during IE execution (IE prior to infection versus after infection and IE prior to infection versus after infection and reboot). Results showed that for both normal and anomalous power profiles, the power consumption was similar. In addition, some delays were seen on the general-purpose computer after it was infected. Figure~\ref{fig:IEPriorAfterInfection12VCPU} showsthe power consumption for opening IE prior to infection versus opening IE after infection with Alureon 

After analyzing all six datasets (three datasets per rootkit), we concluded that a distinguishable difference cannot be made by the naked eye when analyzing the normal and anomalous power profiles for the +12V CPU rails. These results are not the ones we expected because by monitoring the CPU of the general-purpose computer we thought these rails would be more informative. However, we are aware that many processes are running and this extra work consumes more power making it difficult to establish a difference by the naked eye. However, it is possible that the normal and anomalous power profiles may be distinguished by using machine learning algorithms.


\section{Conclusions}
\label{sec:conclusions}
In this paper we presented a proof of concept whose objective was to investigate whether malware leaves a signal on the power consumption of the general-purpose computer. Power data was collected for four rails (+3.3V, +5V, +12V, and +12V CPU) in two different states (normal and anomalous) for two different rootkits. A comparison between the power consumption of the normal and anomalous state was made for each of the events that were recorded. 

The results showed that malware undoubtedly leaves a detectable signal on the power consumption of a general-purpose computer. The signal on the +12V rails on the motherboard was the most useful when identifying an increment in the power consumption after the machine was infected. Results for Alureon showed that when the general-purpose computer was idle (idle prior to infection versus idle after infection and idle prior to infection versus idle after infection and reboot) in a 66.67\% of the time an increment in the power profiles was noticeable by the naked eye, while for Pihar this increment in power was seen in 100\% of the time. For both Alureon and Pihar, there was a 33.33\% of the time in which a notable power signal was seen after the Alureon infection when IE was opened (IE prior to infection versus IE after infection and IE prior to infection versus IE after infection and reboot). In the case of Pihar, 33.33\% of the time an increment was noticeable in the power after infection when opening IE prior to infection versus after infection. When comparing IE prior to infection versus infection and reboot with Pihar, we noticed an increment in the power consumption 100\% of the time.

Besides the +12V rails, the +5V rails are also a valuable parameter to obtain an increment in the power consumption after infection. Results for Alureon showed that 66.67\% of the time there was an increment in the power consumption when comparing idle prior to infection versus idle after infection and when comparing IE prior to infection versus after infection. In the case of Pihar, a noticeable increment in the power consumption was seen 100\% of the time when comparing idle prior to infection versus idle after infection. However, when comparing opening IE prior to infection versus after infection with Pihar we noticed an increment in the power consumption after infection only 33.33\% of the time.

While we obtained promising results, more rootkit samples and complex data analytics are needed to test and validate this approach. In addition, while all the processes running on the machine consumes power, distinguishing between  the normal and anomalous behavior for the general-purpose computer is a challenge because this device is not limited to a certain amount of instructions. Increasing in this way the false positives. 

As part of our future work we intend to include more rootkit samples and workloads in the experimental design and data collection process. Furthermore, we plan to propose an approach that can minimize the number for false positives. In addition, we plan to incorporate machine learning techniques to automatically distinguish between the normal and anomalous power profiles and detect malware.




\section*{Acknowledgments}
Research sponsored by the Laboratory Directed Research and Development Program of Oak Ridge National Laboratory, managed by UT-Battelle, LLC, for the U. S. Department of Energy. This material is based upon work supported by the U.S. Department of Energy, Office of Energy Efficiency and Renewable Energy, Building Technologies Office. Katerina Goseva-Popstojanova's work was funded in part by the NSF under grant CNS-1618629. The authors thank Darren Loposser from the Research Instrumentation group at ORNL for his contributions to this project by providing electronics and sensor support. 


\bibliographystyle{IEEEtran}
\bibliography{IEEEabrv,refs}

\end{document}